\documentclass[%
 aip,
 rsi,
 amsmath,amssymb,
 reprint,%
floatfix
]{revtex4-1}

\usepackage{graphicx}
\usepackage{dcolumn}
\usepackage{bm}

\usepackage[utf8]{inputenc}
\usepackage[T1]{fontenc}
\usepackage{mathptmx}
\usepackage{color}
\usepackage{siunitx}
\usepackage{gensymb}
\usepackage{soul}

\begin{document}

\preprint{AIP/123-QED}

\title[CRESUSOL]{A new instrument for kinetics and branching ratio studies of gas phase collisional processes at very low temperatures
}

\author{O. Durif}
\affiliation{Univ Rennes, CNRS, IPR (Institut de Physique de Rennes) - UMR 6251, F-35000 Rennes, France}

\author{M. Capron}
\affiliation{Univ Rennes, CNRS, IPR (Institut de Physique de Rennes) - UMR 6251, F-35000 Rennes, France}

\author{J. P. Messinger}
\affiliation{Division of Chemistry and Chemical Engineering, California Institute of Technology, Pasadena, California 91125, United States}

\author{A. Benidar}
\affiliation{Univ Rennes, CNRS, IPR (Institut de Physique de Rennes) - UMR 6251, F-35000 Rennes, France}

\author{L. Biennier}
\affiliation{Univ Rennes, CNRS, IPR (Institut de Physique de Rennes) - UMR 6251, F-35000 Rennes, France}

\author{J. Bourgalais}
\affiliation{LATMOS/IPSL, UVSQ, Université Paris-Saclay, UPMC, Univ Paris 06, CNRS, 78280 Guyancourt, France}
\affiliation{Centre de Recherches Pétrographiques et Géochimiques, UMR 7358 CNRS—Université de Lorraine, 15 rue Notre Dame des Pauvres, BP 20, F-54501 Vandoeuvre-lès-Nancy, France}

\author{A. Canosa}
\affiliation{Univ Rennes, CNRS, IPR (Institut de Physique de Rennes) - UMR 6251, F-35000 Rennes, France}

\author{J. Courbe}
\affiliation{Univ Rennes, CNRS, IPR (Institut de Physique de Rennes) - UMR 6251, F-35000 Rennes, France}

\author{G.A. Garcia}
\affiliation{Synchrotron SOLEIL, L’orme des merisiers, BP48 St Aubin, 91192 - Gif Sur Yvette Cedex, France}

\author{J.F. Gil}
\affiliation{Synchrotron SOLEIL, L’orme des merisiers, BP48 St Aubin, 91192 - Gif Sur Yvette Cedex, France}

\author{L. Nahon}
\affiliation{Synchrotron SOLEIL, L’orme des merisiers, BP48 St Aubin, 91192 - Gif Sur Yvette Cedex, France}

\author{M. Okumura}
\affiliation{Division of Chemistry and Chemical Engineering, California Institute of Technology, Pasadena, California 91125, United States}

\author{L. Rutkowski}
\affiliation{Univ Rennes, CNRS, IPR (Institut de Physique de Rennes) - UMR 6251, F-35000 Rennes, France}

\author{I.R. Sims}
\affiliation{Univ Rennes, CNRS, IPR (Institut de Physique de Rennes) - UMR 6251, F-35000 Rennes, France}

\author{J. Thi\'{e}vin}
\affiliation{Univ Rennes, CNRS, IPR (Institut de Physique de Rennes) - UMR 6251, F-35000 Rennes, France}

\author{S.D. Le Picard}\email{sebastien.le-picard@univ-rennes1.fr}
\affiliation{Univ Rennes, CNRS, IPR (Institut de Physique de Rennes) - UMR 6251, F-35000 Rennes, France}

\date{\today}

\begin{abstract}
A new instrument dedicated to the kinetic study of low-temperature gas phase neutral-neutral reactions, including clustering processes, is presented. It combines a supersonic flow reactor with Vacuum Ultra-Violet (VUV) synchrotron photoionization time of flight mass spectrometry. A photoion-photoelectron coincidence detection scheme has been adopted to optimize the particle counting efficiency. The characteristics of the instrument are detailed along with its capabilities illustrated through a few results obtained at low temperatures (< 100 K) including a {photoionization spectrum} of n-butane, the detection of formic acid dimer formation as well as the observation of diacetylene molecules formed by the reaction between the C$_2$H radical and C$_2$H$_2$.  
\end{abstract}

\maketitle




\section{Introduction}

Understanding the mechanisms of elementary reactions leading to the formation of molecules and clusters in various conditions of temperature and pressure, is of fundamental interest and yields crucial information for modeling gaseous environments encountered in the fields of combustion, atmospheric chemistry and astrochemistry.\citep{Seakins_2007,Osborn_2017,Cooke_2019}  Both rate coefficients and quantitative information on the reaction products of elementary collisions are important pieces of information for improving the accuracy of chemical networks involved in planetary and interstellar photochemical models.  In particular, the proportion of each accessible product channel of the chemical reactions involved in these models are essentially missing at the low temperatures prevailing in many astrophysical objects such as dense interstellar clouds or cold planetary atmospheres.
This lack of data can have dramatic effects on astrochemical networks, especially when exit channels are missing as reaction products may further react. The knowledge of reaction products and their branching ratios also provides valuable information to infer the mechanism of elementary reactions and is therefore of fundamental interest. 
Certain quantum effects such as tunneling may be enhanced at low temperatures and their proper inclusion in theoretical calculations remains a challenge.\citep{shannon2010, shannon2013, sims2013,tizniti2014,shannon2014,gomez2014,caravan2015, antinolo2016, jimenez2016, ocana2018, heard2018,ocana2019, blazquez2019} Experiments performed at low temperature are therefore essential to benchmark theoretical approaches and provide reliable data for chemical modeling. \citep{LePicard_2019} 

In order to identify and quantify reaction products, the experimental methods used have first to be highly sensitive, as some products can be formed at very low yields, potentially in a large variety of internal states. They should also be able to detect a wide range of different products at the same time with sufficient time resolution to enable elementary reaction kinetics to be followed in real time for several products and/or reactants,  all of which may have potentially very different abundances. In view of this, the dynamic range of detection needs also to be high.
Diagnostic methods have  significantly improved in the last two decades and most experimental approaches involve optical and microwave/millimeter spectroscopy or mass spectrometry techniques. The main drawback of optical techniques is that it is difficult to monitor a large range of species with one single apparatus. In some favorable cases laser induced fluorescence (LIF) can be applied to determine the H atom product yields of a reaction by monitoring the H atom production.\citep{Seakins_2007} This quantitative information requires the use of a calibration reaction and does not provide any quantitative information on H atom channel co-product(s) or other reaction channels. In most cases, this technique needs therefore to be completed by other methods.\citep{Bourgalais_2015} Recently, a pulsed supersonic uniform flow has been combined with a cw-CRDS {(Cavity Ring-Down Spectrometer)}  operating in the near infrared for spectroscopy and kinetics at low temperature. \cite{Suas-David_2019} This new high resolution and sensitive absorption spectrometer offers the possibility to probe numerous species which have not been investigated yet. It has been designed to probe radicals and reaction intermediates and also to follow the chemistry of hydrocarbon chains and PAHs.  New developments in time-resolved frequency comb spectroscopy,\citep{Schliesser_2005} with broadband and high resolution capabilities in the mid-infrared,\citep{Adler_2010, fleisher_2014, bjork_2016, Bui_2018} appear to be also very promising as this technique possesses the advantage of being sensitive and allowing detection of multiple species simultaneously. Provided they have a permanent dipole, reaction products can also be detected using rotational spectroscopy, which has the advantage of being highly specific. It has traditionally suffered however, from a lack of sensitivity compared to laser based techniques that has prevented its use in reaction kinetics and dynamics. Recently, the Chirped Pulse Fourier-Transform MicroWave (CPFTMW) spectroscopic technique developed by Pate and co-workers has opened new perspectives by improving the rate of data acquisition by several orders of magnitude, as well as by covering a wide range of frequencies that enables simultaneous detection of multiple species.\citep{brown_broadband_2008,park_perspective:_2016} This technique has been recently coupled to a CRESU {(Cinétique de Réaction en Ecoulement Supersonique Uniforme or Reaction Kinetics in Uniform Supersonic Flow)} apparatus, to study gas phase reaction kinetics in a very low temperature environment (down to 20 K). Referred to as the CPUF (Chirped Pulse in Uniform Flow) technique, it has demonstrated its capability to study bimolecular reaction products.\citep{oldham_chirped-pulse_2014, abeysekera_chirped-pulse_2014} 
Photoionization /mass spectrometry (PIMS) techniques have the advantage of being both universal and sensitive in contrast to the techniques mentioned above, as any molecule can be ionized in the VUV with a high cross-section and ions counted individually, with no background. 
It is particularly powerful when coupled to synchrotron radiation as the photon energy tunability allows threshold photoionization, limiting fragmentation of the detected species, so that the parent cation can be monitored. The tunability of the synchrotron can, under certain conditions, give access to the isomeric structures of the products through their photoionization spectra. Branching ratios are obtained by fitting photoionization data with experimental or calculated photoionization spectra of the pure compounds. \citep{osborn_multiplexed_2008} Great success has been achieved by  Taatjes, Osborn, and coworkers performing experiments at room temperature or above for atmospheric and combustion applications,\citep{Osborn_2017} and by Qi and coworkers for combustion and catalysis studies. \citep{qi_combustion_2013,jiao_selective_2016} A few low-temperature experiments have also been performed at the ALS (Advanced Light Source) down to \SI{70}{\kelvin} using a pulsed CRESU apparatus coupled to a quadrupole mass spectrometer,\citep{Lee_2000,Soorkia_2010,Soorkia_2011,Bouwman_2012, Bouwman_2013} {primarily} for reactions of interest for the atmosphere of Titan, the largest satellite of Saturn. 

Here we present a new experimental apparatus, CRESUSOL, coupling the CRESU technique to a photoionization time of flight mass spectrometer using the synchrotron radiation at SOLEIL (France). This setup is dedicated to the monitoring of reactants and products of neutral-neutral reactions at substantially lower temperatures than has been attempted before (down to 50 K). Its goal is to measure the rate coefficients of neutral-neutral reactions, including dimerization, as well as their branching ratios. The uniform supersonic flow reactor based on the CRESU technique is described in Section II while the photoionization mass spectrometer chamber along with the synchrotron source, the time of flight mass spectrometer and the data acquisition set up, are outlined in Section III. Section IV presents a {photoionization spectrum} of n-butane at 132 K, the detection of formic acid dimer formation, as well as the observation of diacetylene molecules formed by the reaction between the radical C$_2$H and C$_2$H$_2$.  The photoionization spectrum of n-butane and the detection of diacetylene, C$_4$H$_2$, produced by the reaction between C$_2$H and C$_2$H$_2$ at 50 and 70 K are presented as proof-of-principle results. Concluding remarks and perspectives are given in Section V.  




\section{Uniform {supersonic} flow chemical reactor}
\label{sec_reaction_chamber}

\begin{figure}[h]
 \includegraphics[width=\columnwidth]{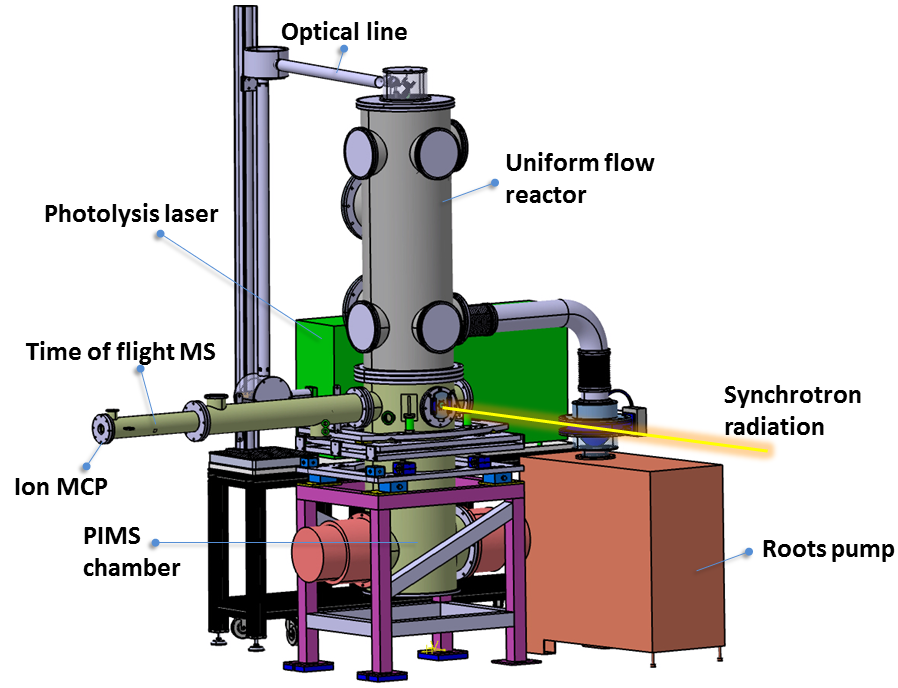}
 \caption{Outside view of the whole setup (see text for details).}
 \label{fig:setup_whole}
\end{figure}

The CRESU technique was developed in the early eighties at the Laboratoire d'A\'erothermique de Meudon in France by Rowe, Marquette and their colleagues, \citep{dupeyrat_design_1985} and since then it has been described many times.\citep{potapov_2017,smith_gas_2008}
Hence, only a brief overview will be reported here with particular attention given to specific aspects inherent to the present work. The strength of the technique is the ability to generate routinely cold flows of gases at temperatures as low as \SI{20}{\kelvin} and even below,\citep{Berteloite_2010} while maintaining them in the gas phase under supersaturation conditions. The CRESU is a wall-less reactor consisting of a supersonic flow obtained by the continuous isentropic expansion of a buffer gas from a reservoir to a low pressure chamber through a suitably designed convergent-divergent Laval nozzle. \citep{dupeyrat_design_1985, jimenez_development_2015} The uniformity of the flow in terms of velocity, temperature, pressure and hence density ensures that {these properties are well defined for typically a few tens of centimeters parallel to the flow direction, and for about one centimeter perpendicular to the flow direction in the present apparatus. This corresponds to durations of a few hundreds of microseconds on the flow axis.} This time is significantly shorter than those accessible in standard reaction cells, which prevents the measurement of slow processes, i.e. with rate coefficients smaller than 10$^{-12}$ cm$^3$ s$^{-1}$. Although the reactant density could be increased to compensate for this short time, it must be kept small enough with respect to the main buffer gas density in order to prevent the loss of uniformity of the supersonic flow.  The concentration for the reactants should not exceed typically \SI{1}{\percent} of the total density of the main flow. Typical mass flow rates ranging from 10 to 100 standard liters per minute (slm) are required to generate uniform supersonic flows with a density of a few 10$^{16}$ cm$^{-3}$, necessitating pumping capacities of the order of  \SI[per-mode=symbol]{10 000} - \SI[per-mode=symbol]{20 000} {\m\cubed\per\hour}. Larger pumping capacities are necessary when the temperature needs to be lowered (implying higher Mach number flows) and/or when the pressure of the uniform flow needs to be decreased. This makes such CRESU reactors large, heavy, expensive and not easily transportable. 

In order to couple a CRESU device to the synchrotron radiation at SOLEIL (France), a dedicated apparatus had to be designed, more compact and with reduced pumping capacities to fit as a non-permanent endstation on the VUV beamline DESIRS. Such a reduction of pumping speed prevents the generation of continuous supersonic flows at temperatures lower than typically 50 K. To circumvent this disadvantage, pulsed uniform supersonic flows will be implemented in the longer term. Here, the proof of concept of the CRESUSOL apparatus is demonstrated using a continuous supersonic flow at intermediate temperatures, at i.e. \SI{50}{\kelvin} and above as a major first step. 

\begin{figure}[h]
 \includegraphics[width=\columnwidth]{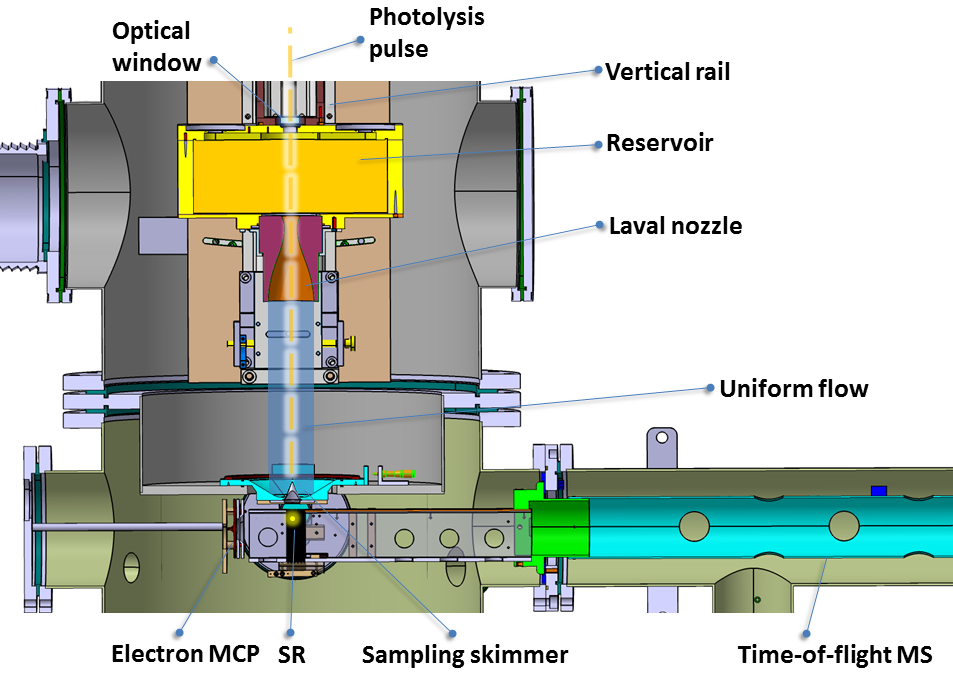}
 \caption{Inside view of the reaction and detection chambers. The ion micro-channel plate detector, implemented at the end of the drift tube, is out of {view}.}
 \label{Movable_nozzle}
\end{figure}

\begin{figure}[h]
 \includegraphics[scale=0.50]{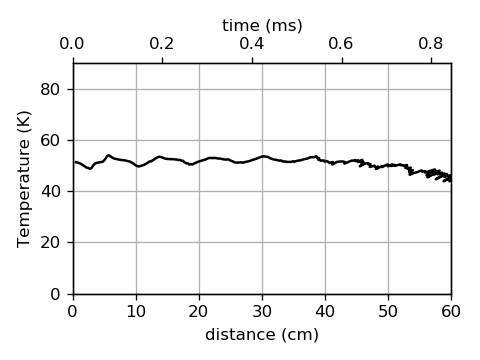}
 \caption{Axial temperature profile for the Laval nozzle operating at 52 K (see table \ref{table:Nozzle}) with N$_2$ as the buffer gas, determined by impact pressure measurements using a Pitot tube. The uniform zone of \SI{\sim40}{cm}  gives an average value for the temperature of (52.0 $\pm$ 1.1) K. Downstream of the uniform zone the flow is no longer isentropic, and the relation linking the temperature to the impact pressure breaks down. The temperature is actually higher in this part of the flow.}
 \label{fig:pitot_profil}
\end{figure}

For this purpose a \SI{2.75}{-m} long, \SI{0.5}{-m} diameter chamber has been designed in two sections (see Figs. \ref{fig:setup_whole}--\ref{Movable_nozzle}). The upper one is a \SI{1.5}{-m} long CRESU chamber where supersonic flows are generated using Laval nozzles mounted onto a reservoir (10 L). It is pumped through a Roots blower (Edwards GXS 450/4200) with a peak pumping speed of \SI[per-mode=symbol]{3026}{\m\cubed\per\hour} for molecular nitrogen. Several dedicated Laval nozzles with flow temperatures ranging from \SI{\sim 50}{\kelvin} to \SI{\sim 130}{\kelvin} have been specifically characterized (see table \ref{table:Nozzle}). The supersonic flows generated by these Laval nozzles were probed using a Pitot tube.\citep{sims_1994}  Briefly, when placed at the center of the supersonic flow the Pitot tube causes a detached shock wave upstream of its extremity. The impact pressure of this shock wave measured via the Pitot tube is directly related to the Mach number $\mathcal{M}$, of the supersonic flow, its temperature and its density. The uniformity of the flow is then scanned by varying the distance between the extremity of the Pitot tube and the exit of the nozzle. Figure \ref{fig:pitot_profil} shows the temperature profile recorded with this technique for a flow generated by a Laval nozzle operating at 52 K, designed for this first series of experiments.

\begin{table}
\begin{center}
\bgroup
\def\arraystretch{1.3}
{\tabcolsep=3.5pt
\begin{tabular}{llllll}
\hline 
\hline  
$T$ /K & Gas &  $n /{10^{16}}$cm$^{-3}$ & $P_{\rm{ch}}$ /mbar & $\mathcal{M}$ & $t_{\rm max}$/$\mu$s \\ 
\hline
$52.0 \pm 1.1$ & N$_2$ & $3.8 \pm 0.2$ & $0.28$ & $4.8 \pm 0.1$ & $570$ \\ 
$74.6 \pm 3.4$ & N$_2$ &$6.6 \pm 0.8$ & $0.75$ & $3.8 \pm 0.1$ & $640$ \\ 
$97.6 \pm 1.6$ & N$_2$ & $7.6 \pm 0.3$ & $1.13$ & $3.2 \pm 0.1$ & $360$\\ 
$98.8 \pm 1.2$ & N$_2$ & $4.2 \pm 0.1$ & $0.63$ & $4.2 \pm 0.1$ & $390$ \\ 
$132 \pm 2$ & Ar & $3.4 \pm 0.1$ & $0.62$ & $ 3.5 \pm 0.1$ &  $250$ \\ 
\hline
\hline
\end{tabular}
}
\egroup
\end{center}
\caption{Experimental conditions of the supersonic uniform flows generated by a set of Laval nozzles. $t_{\rm max}$ corresponds to the time during which the supersonic flow remains uniform, $\mathcal{M}$ is the Mach number  .}
\label{table:Nozzle}
\end{table}




\section{Photoionization Mass Spectrometry (PIMS) chamber}

The detection chamber (or PIMS chamber) is hosted in a 1.25-m long, 0.5-m diameter stainless steel cylinder situated below the CRESU chamber (see Fig. \ref{Movable_nozzle}). It is equipped with a uniform flow sampler (Sec. \ref{subsection_sampling}) implemented on the top flange. The introduced gases are analyzed by a 2-m long time-of-flight mass spectrometer (Sec. \ref{subsection_PIMS}) coupled to the VUV synchrotron radiation for photoionization (Sec. \ref{subsection_SR}). A vacuum of $10^{-6}$ mbar is maintained during operation by two magnetically levitated turbomolecular pumps (Edwards STP-IX3006C) with a combined capacity of 5 300  L s$^{-1}$ for N$_2$ and backed by two dry scroll pumps (Edwards nXDS20i). 
{This relatively large pumping capacity is necessary to ensure a sufficiently low pressure ($\leq$ 10$^{-5}$ mbar) in the PIMS chamber while sampling the continuous supersonic flow in order to avoid damage to the ion and electron MCPs (micro-channel plates) and to prevent contamination of the DESIRS beamline.}

The vacuum in the free drift tube of the mass spectrometer is further improved by the addition onto the tube of two turbomolecular pumps (Edwards nEXT 300D) with a total capacity of 300 L s$^{-1}$ for N$_2$. The PIMS chamber is mounted onto a metal frame. The frame design allows vertical movement of the PIMS chamber over several centimeters via several jacks while rotational and translational degrees of freedom are ensured by a system of ball bearing plates. These adjustments allow the setup to be finely aligned in vacuum with the synchrotron radiation beam which must be maintained parallel to the electrode plates of the mass spectrometer and must intersect the center of the molecular beam (Sec. \ref{subsection_molecular_beam}) generated by the sampling system.

\subsection{Uniform supersonic flow sampling}
\label{subsection_sampling}

The relatively dense (10$^{16}$ $-$ 10$^{17}$ cm$^{-3}$) uniform flow cannot be probed directly by mass spectrometry as long as low pressure conditions required for good operation of the detectors are not fulfilled. 
To reduce the pressure by 5 to 6 orders of magnitude, the flow is sampled by a skimmer mounted onto a streamlined flange separating the reaction chamber, described in Sec. \ref{sec_reaction_chamber}, from the PIMS chamber. The sampling arrangement (see Fig. \ref{Movable_nozzle}) leads to the generation, from the dense uniform supersonic flow, of a molecular beam, basically free of collisions i.e. freezing any downstream reactions. 

Although considered as minimally intrusive, the insertion of a skimmer in the flow can induce some perturbations and affect the density and temperature of the gas in the probed region so a careful design is therefore required. To that end, Computational Fluid Dynamics (CFD) calculations were performed. The critical features in the sampling of the supersonic uniform expansion were found to be the shape of the skimmer and the distance from the apex of the skimmer to the flat wall orthogonal to the flow direction on which the skimmer is mounted. 

\begin{figure}[h]
 \includegraphics[width=\columnwidth]{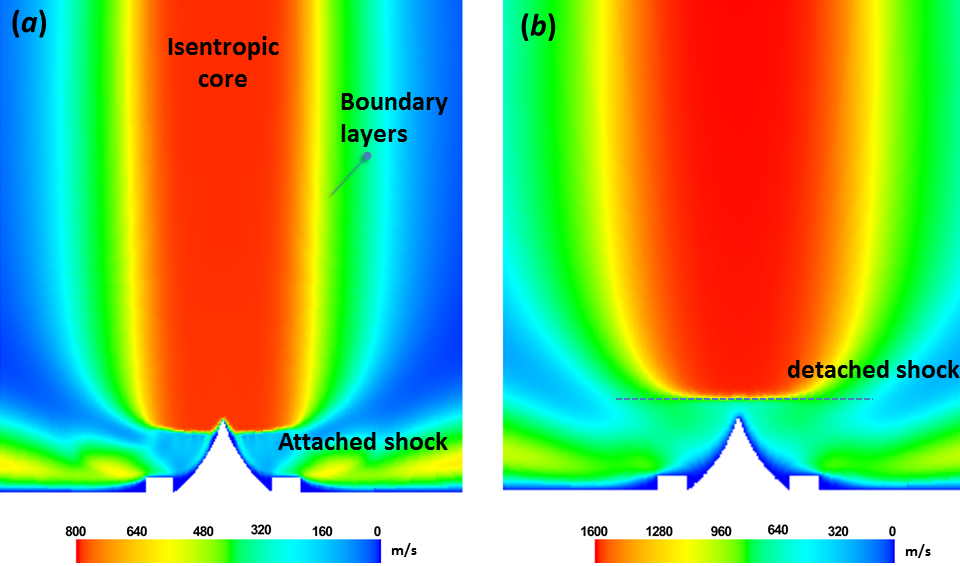}
 \caption{Velocity color maps -- obtained by CFD simulations -- of uniform flows sampled by a 25-mm long skimmer mounted on a flat flange. ($a$) Mach 3.8 uniform expansion of N$_2$ at $T=$75 K and with a density of 1.7$\times 10^{16}$~cm$^{-3}$. The shock wave remains attached to the tip of the skimmer and the cold isentropic core is sampled. ($b$) Mach 4.6 uniform expansion of He at $T=$36 K with a density of  5.3$\times 10^{16}$~cm$^{-3}$. A detached normal shock wave is produced by the expansion of helium. Under these conditions, the warm gas behind the shock wave, and not the isentropic core, is sampled.}
 \label{Fig_detached_layer}
\end{figure}

To minimize the gas throughput, i.e. to keep the pressure in the PIMS chamber $\leq 10^{-5}$~mbar, a small skimmer orifice diameter was selected. The shape of the skimmer, chosen among the commercially available models  (Beam Dynamics), is characterized by an orifice diameter of 0.1 mm, an orifice edge thickness of 10 $\mu$m, and a total included angle at orifice of 25$\degree$ internal and 30$\degree$ external. Calculations were initially carried out for the two skimmer lengths of 15 and 25 mm. Different types of carrier gas (Ar, He and N$_2$) and different flow temperature and density conditions were tested in the simulations to highlight the disturbances caused by the interaction between the expansion and the sampling probe system. 

Results from initial CFD calculations, a sample of which is displayed in Fig. \ref{Fig_detached_layer}, show that for a flat separating flange, detachment of the shock wave from the surface occurs but more importantly can occur as well in front of the skimmer, thereby heating the gas before its introduction into the PIMS chamber. This configuration is not adapted to kinetic measurements, in particular when van der Waals complexes are present in the flow which may break up in this warm layer. While the isentropic core of a Mach 3.8 N$_2$ flow at $T=75$ K  can be sampled, this is not the case for the colder and faster Mach 4.6  He flow at $T=36$ K for which the normal shock layer is clearly detached (see Fig. \ref{Fig_detached_layer}(b)).

The position of the shock wave is governed by inertial forces, which depends on the Mach number $\mathcal{M}$, the total mass of the gas and the heat capacity ratio $\gamma$, as well as the angle of attack $\alpha$. The presence of a flat obstacle in the flow such as the support flange of the skimmer ($\alpha=90\degree$), is identified as the cause of the detachment of the shock wave. By contrast, the skimmer with its slender shape plays no significant role in the localization of the shock wave. For a given gas, the faster the flow (large $\mathcal{M}$), the smaller the compression zone and the closer the shock wave from the obstacle. In other words a faster flow increases the thrust which compresses more gas at the obstacle and therefore reduces this 
area.

\begin{figure}[h]
 \includegraphics[width=\columnwidth]{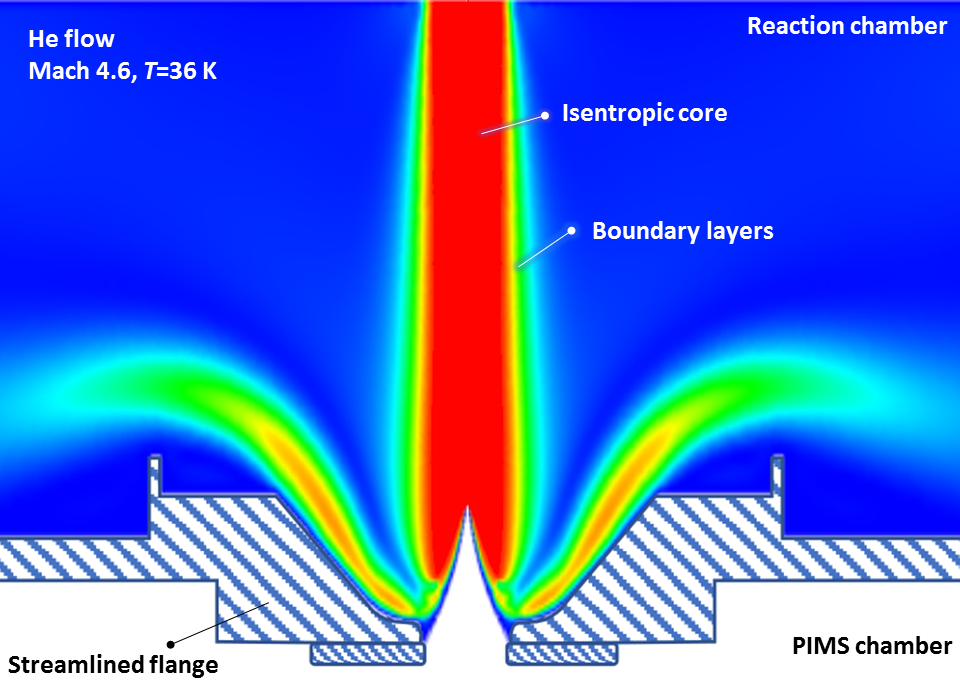}
 \caption{Velocity color map of a Mach 4.6 He flow sampled by a 25-mm long skimmer mounted onto a streamlined separating wall. The shock remains fully attached to the skimmer over a distance about half of its length, ensuring a sampling of the cold ($T=$36~K)  isentropic core of the uniform flow.}
 \label{fig:sampling_wall}
\end{figure}

To reattach the layer, a streamlined separating wall was designed with the help of CFD calculations. As shown in Fig. \ref{fig:sampling_wall} for a 36 K He flow, the presence of a streamlined flange induces the disappearance of the detached layer by facilitating the residual gas evacuation. This configuration is much more universal and gives us access to the whole set of available Laval nozzles displayed in Table \ref{table:Nozzle}.

\subsection{Molecular beam properties}
\label{subsection_molecular_beam}

As the divergence of the molecular beam causes its central density to drop quickly, the interface between the reaction and PIMS chambers was also designed to minimize the distance between the skimmer apex and the photoionization region. It has also been shown that this is important in order to minimize loss of kinetic time resolution induced by the distribution of molecular speeds downstream from the skimmer.\citep{taatjes_2007} The ionization is carried out at \SI{\sim 53}{\mm} from the apex of the skimmer. This distance was kept as small as possible, given the height of the skimmer, the height of the skimmer support and the bulk of the electrodes of the mass spectrometer. To estimate the particle density as a function of the sampling distance, we performed a basic Monte-Carlo simulation of random trajectories with initial velocity distribution of particles given by the flow temperature. Collisions between molecules in the beam, with the residual gas or with the walls of the skimmer, were not taken into account. 
{The simulation shows that less than 1\% of the supersonic flow density remains in the ionization region.}
As the total gas densities of the Laval nozzles fall in the $10^{16} - 10^{17}$ cm$^{-3}$ range, the  beam density is expected within the $10^{14} - 10^{15}$ cm$^{-3}$ range.   

\subsection{Photoionization radiation}
\label{subsection_SR}
Intense and tunable VUV radiation was supplied by the undulator-based beamline DESIRS\citep{nahon_desirs:_2012} at Synchrotron SOLEIL. The present data was obtained by using the low dispersion/low energy 200 grooves/mm grating of the monochromator providing a typical flux of $10^{12} - 10^{13}$ ph/sec for a \SI{0.1}{\percent} bandwith in the \SIrange{5}{20}{\eV} range. The exit slit we used (50 microns) led to a typical photon resolution of \SI{3.5}{\milli\eV} around \SI{11}{eV} and to a FWHM transverse footprint at the focal point, corresponding to the ionization region, of 200 microns (H) x 90 microns (V). The crucial issue of spectral purity was addressed by using a gas filter filled with Ar suppressing by 4 to 5 orders of magnitude any undulator high harmonics that would be transmitted by the grating’s high orders. \citep{mercier_experimental_2000}
Furthermore, the beamline post-focusing last mirror, mounted on a goniometer, can be used to precisely align the radiation with the center of the molecular beam in between the electrodes defining the interaction region of the mass spectrometer. This position fine tuning is relatively critical to optimize the quality of the signal.

\subsection{Time-of-flight mass spectrometer}

\label{subsection_PIMS}
Reactants, introduced at low concentrations in the  supersonic flows, and their subsequent products are probed by a modified Wiley-McLaren time-of-flight mass spectrometer \citep{wiley_timeflight_1955} collecting ions and electrons particles in coincidence. 
Such a configuration provides multiplex detection and 100\% transmission for all the ions, only limited by the detection efficiency of the microchannel plates (about 60\% \cite{krems_2005}), a clear advantage over quadrupole mass analyzers which transmit only the ions with a given {mass-to-charge} ratio and need to be scanned to acquire the whole mass spectrum. This is, however, at the expense of a decreased dynamical range, but multiplex acquisition is crucial in the present type of experiments where the $m/z$ information is recorded as a function of other variables that also need to be scanned (photon energy and reaction time for instance).
Note that presently the electrons only serve as the start of the ion time of flight to avoid pulsing the extraction field. %

\begin{figure}[h]
 \includegraphics[width=\columnwidth]{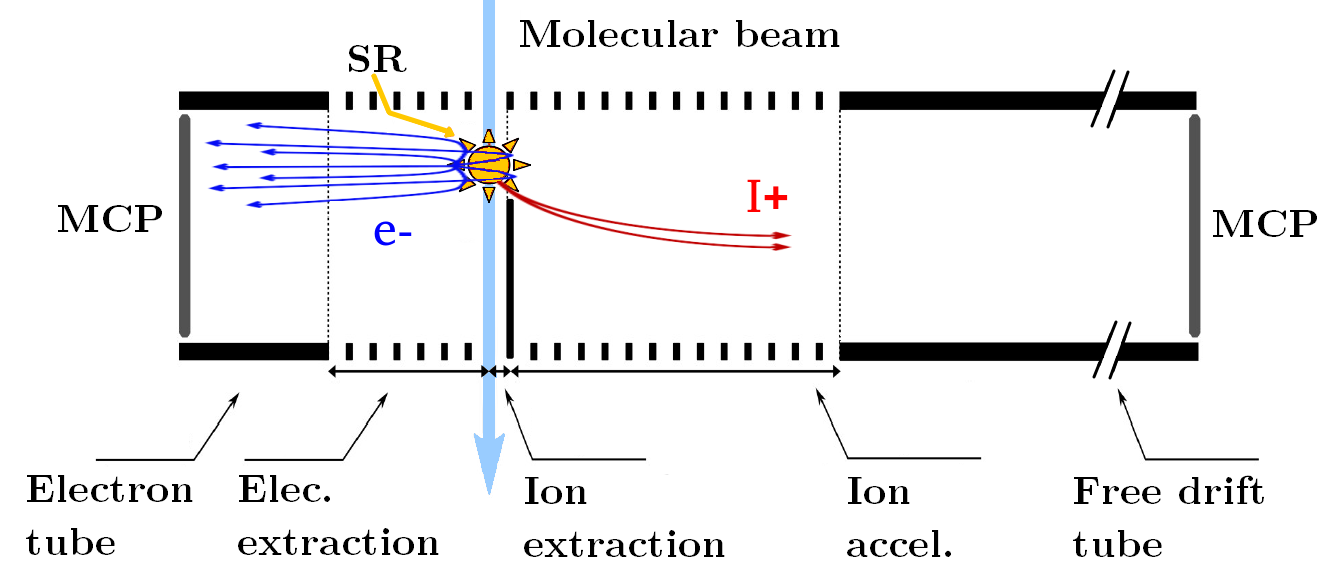}
 \caption{Schematic representation of the CRESUSOL mass spectrometer (not to scale).}
 \label{fig:spectro}
\end{figure}

A velocity map imaging electron and/or ion spectrometer providing complementary spectroscopic information will be implemented in the future. 
Under the current configuration, depicted in Fig. \ref{fig:spectro}, the synchrotron radiation (SR) is centered between the first electrode on the electron side and the electrode on the ion side that are separated by 4.6 mm, and intersects the molecular beam near the upper edge of the extraction region, in order to minimize the distance from the skimmer apex (see  III A.). 

Ions are extracted continuously by static fields through two acceleration regions and guided towards the free flight tube, following the Wiley-McLaren type geometry.\citep{wiley_timeflight_1955} 
{In contrast to the ion extraction region, the field at the end of the electron extraction side is not homogeneous to allow focusing of the more energetic electrons onto the detector.}
In order to obtain the greatest resolving power, $m/\Delta m$, where $m$
is the molecular weight and $\Delta m$ is the corresponding full width
at half maximum (FWHM), we opted to minimize the length of the two acceleration zones and to maximize the free flight tube length $D$ (see Table \ref{table:TOF}). The first acceleration zone (or extraction zone), $s_0$, is 2.3 mm long and lies between the ionization point and the first electrode. The second acceleration zone, $d$, is 13 mm long, standing between this first electrode and the last electrode at the entrance of the free flight tube. Each electrode was machined in molybdenum, using the Institut de Physique de Rennes workshop facilities. They are square shaped,  70 mm x 70 mm, with a thickness of 0.3 mm. The spacing between each electrode is 0.7 mm in order to obtain the most linear acceleration, and high transmission grids are added on the first and the last electrode to reduce penetrating fields. After acceleration, ions fly field-free in a tube over \SI{200}{\cm} long and are then collected via \SI{40}{\mm} microchannel plates (Photonis PS 36204).

On the electron detection side, six identical electrodes are arranged in the same way as for the ions to accelerate the electrons over 5 mm, and a tube of a few centimeters long extends the extraction to the \SI{25}{\mm} microchannel plates (Photonis PS 30305) located about ten of centimeters from the ionization zone. The electron signal is used to provide the start of the time of flight (TOF) of the ions following a multistart/multistop coincidence scheme.\citep{bodi_2007} as detailed in Sec. \ref{PEPICO}.

In order to minimize the false coincidences due to parasitic secondary electrons emitted by scattered VUV radiation, all the electrodes were covered with water-based colloidal graphite coating (Aquadag) and the synchrotron beam emerging from the ionization zone is deflected at \SI{45}{\degree} on the exit flange to reduce reflections of the synchrotron beam photons towards the electrodes. To reduce electronic noise, voltages applied to the electrodes are stabilized and filtered.


Trajectory simulations of charged particles using SIMION software were performed to validate the geometry of the mass spectrometer, considering the lengths of the two acceleration zones, the free drift tube length, the electrode potentials, and especially taking into account the initial velocity of the molecular beam. The simulated average time of flight of ions and the trajectories of the particles were found to be consistent with what was observed experimentally, however the dispersion of flight times for a given mass was not gaussian. Nevertheless, the observed resolving power, $m/\Delta m$ $\sim$ 800, is found to be close to that theoretically expected by the overall resolution of the Wiley - McLaren equation, $\sim$ 900. This satisfactory resolving power avoids the need for a reflectron in most cases even though an higher resolution would be desirable for performing studies of the dynamics of dissociation of metastable molecular Van der Waals clusters.\citep{shinohara_metastable_1991}

\begin{center}
\bgroup
\def\arraystretch{1.3}
{\tabcolsep=6pt
\begin{table}
\begin{tabular}{lllll}
\hline
\hline   
zone &\text{size} & \text{field } & \text{energy } & \text{time} \\ 
&cm & V.cm$^{-1}$ & eV & $\sqrt{m/z}\, \mu$s \\ 

\hline
extract. & $s_0 = 0.23$ & $ E_{s} = 300 $ & $ q s_0 E_{s} = 69 $ & $t_{s} = 0.04$\\
accel. & $d = 1.3$ & $ E_d = 3178 $ & $ q d E_{d} = 4131 $ & $t_d = 0.026$\\
drift& $D = 200$ & $E_D = 0$ & $U_t = 4200 $ & $t_D = 2.23$\\
 \hline
 \hline
\end{tabular}
\caption{Time-of-flight mass spectrometer specifications. {The terminology is taken from Wiley and McLaren. \cite{wiley_timeflight_1955}}}
\label{table:TOF}
\end{table}
}
\egroup
\end{center}
\subsection{Photoion-photoelectron acquisition}
\label{PEPICO}
\begin{figure}[h]
 \includegraphics[width=\columnwidth]{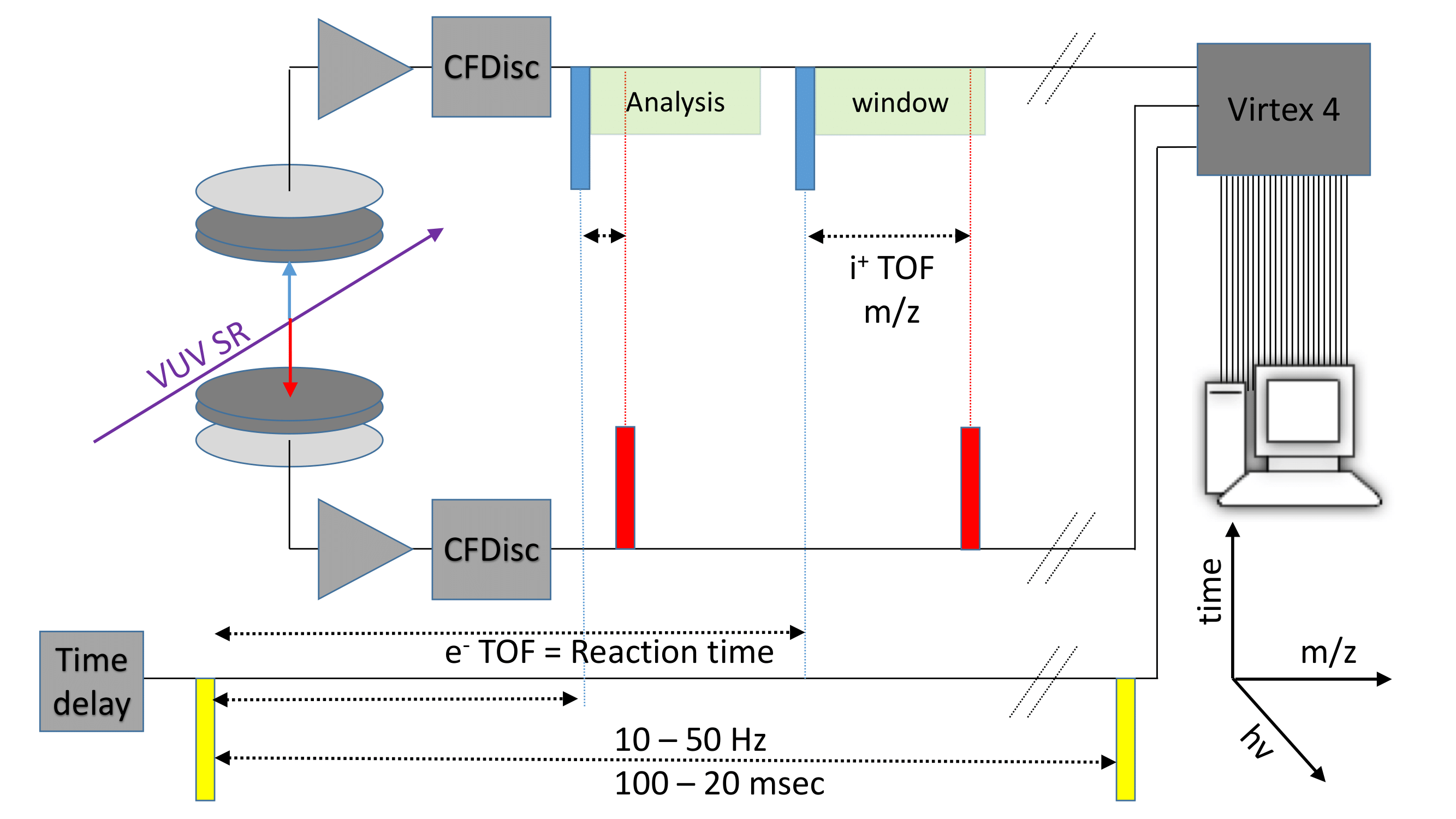}
 \caption{Schematics of the detection setup and timing. Yellow signals represent the periodic excimer laser trigger, while blue and red bars represent the continuous arrival of electron and ion signals resulting from synchrotron photoionization. Ions within the analysis window are considered correlated to the electron that triggers the window.}
 \label{fig:acquisition_scheme}
\end{figure}

The detection scheme is shown in Fig. \ref{fig:acquisition_scheme}. The signals from the back of the MCPs are pre-amplified (FAMP1+, Roentdek) and discriminated via a constant fraction discriminator (Philips), before being fed to a time-to-digital (TDC) converter. The TDC was built in-house within the LUMAT (Lumière Matière) platform in Orsay, France, and is based on a Virtex 4 FPGA (Field-Programmable Gate Array) from Xilinx. In its basic mode, the TDC is capable of sampling flight times of up to 16 ms in steps of 120 ps. The time-of-flight of the photoions is continuously measured relative to that of the photoelectrons using a multistart/multistop coincidence scheme \citep{bodi_2007} that has the advantage of providing an unstructured background of false coincidences which is easily subtracted. In addition, the lack of pulsed fields increases the ultimate mass resolution and the sampling step of the reaction time in photolysis experiments, which is now only limited by the TDC. In this scheme, all ions that arrive within the analysis window that starts with the arrival of the electron are correlated to said electron. The analysis window is defined in the data treatment as the flight time of the heaviest ion expected. For high counting rates and/or {high false} coincidence rates, there will be a probability that uncorrelated ions also arrive within this window. For instance, the {false} coincidence rate can be {increased} by the photoemission effect (photons impacting on metal surfaces in the ionization chamber, yielding electrons but no ions), or by electronic noise. These uncorrelated signals are referred to as false coincidences, and in the multistart/multistop scheme appear as a flat baseline. This baseline needs to be kept to a minimum to avoid increasing the error bars on the background subtraction procedure, thus lowering the detection limit. Because all the species in the ionization region contribute proportionally to the false coincidence baseline, this is even more critical if the relative abundance of the species of interest is low.
For experiments where the reaction time is given by a pulsed photolysis laser, a third, periodic, signal from the laser trigger enters the TDC. Because this signal gives the start of the reaction, the kinetics information is contained in the time difference between the photoelectron and the laser trigger arrival times, as shown in Figure \ref{fig:acquisition_scheme}.  The signal from the laser can be delayed with respect to the actual light pulse in order to adjust the 16 ms acquisition time window offered by the TDC. Note that the photoelectron flight time adds a positive {offset} to the reaction time and a negative {offset} to the ion flight time. However, this {offset} is of the order of a few ns and has virtually no dependence on the electron initial kinetic energy, so that it can be considered constant and negligible both with respect to the TOF of the ions (a few $\mu$s), and to the kinetic times studied with this setup.

Treatment of the timing data yields the number of photoionization events as a function of ion $m/z$, reaction time, and photon energy. Relevant information in this 3D hypermatrix can then be projected and plotted in a 2D intensity map or 1D curve along other selected dimensions, providing highly multiplex information.






\section{Experimental results}
In this section we highlight the performance of the instrument and its capabilities to detect products of reactions including dimerization processes as the proof of concept of the CRESUSOL apparatus in its present configuration. 

\subsection{Butane {photoionization spectrum}}
\begin{figure}[h]
 \includegraphics[width=\columnwidth]{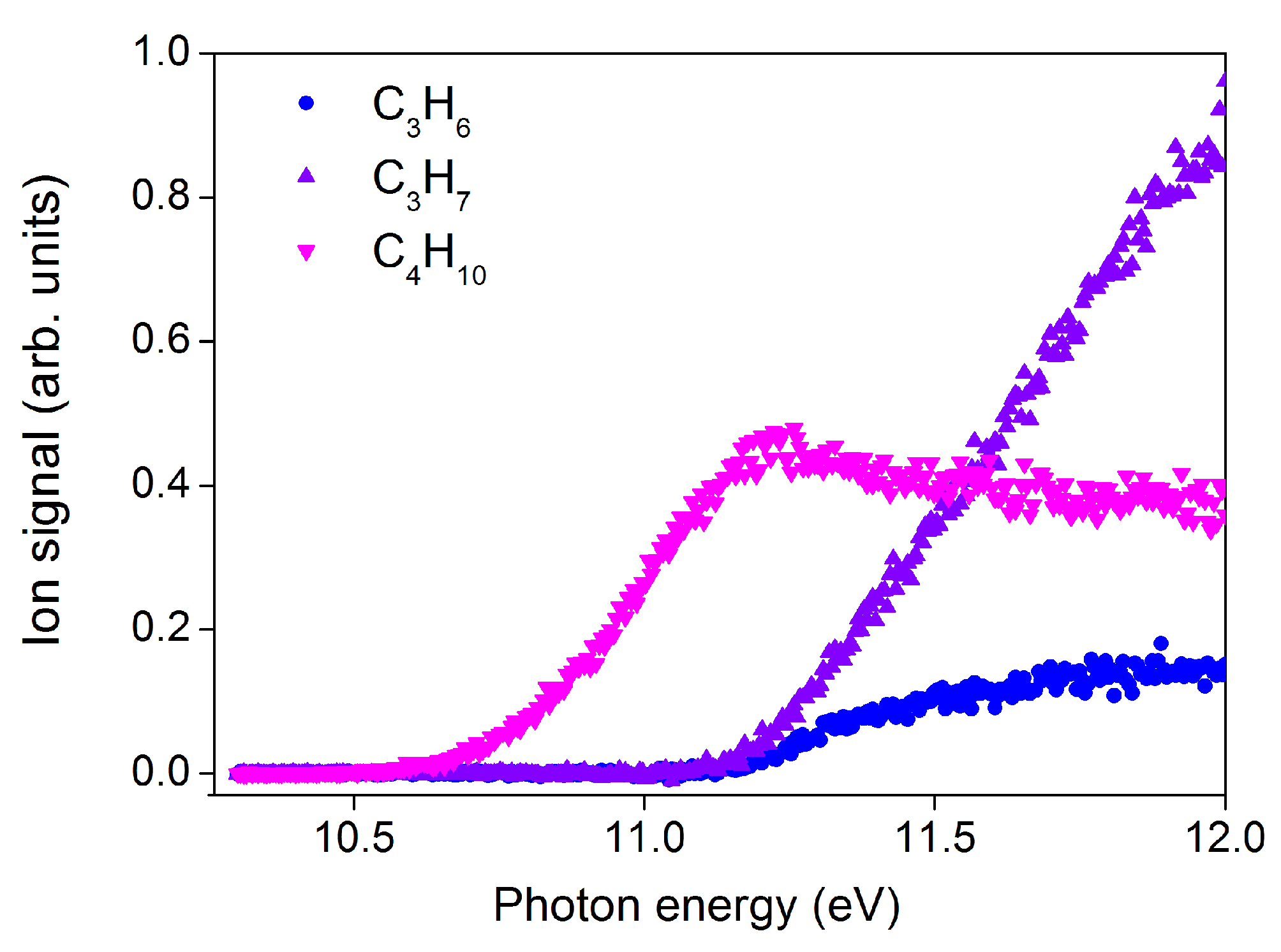}
 \caption{{Photoionization spectrum} of butane ($m/z=58.12$) introduced with a concentration of 5.2 $\times$ 10$^{13}$ cm$^{-3}$ in a supersonic flow operating at 132 K with a buffer gas of argon, {$n_{\text{Ar}}$} = 3.39 $\times$ 10$^{16}$ cm$^{-3}$.}
 \label{butane}
\end{figure}

Acquisition of a photoionization spectrum of a stable species, n-butane, was chosen as a first step for testing the apparatus. 
A small amount of n-butane ($n_{{\text{C}}_4{\text{H}}_{10}}$ = 5.2 $\times$ 10$^{13}$ cm$^{-3}$) was introduced in the  supersonic flow of a Laval nozzle operating at 132 K with argon as the buffer gas ($n_{\text{Ar}}$ = 3.39 $\times$ 10$^{16}$ cm$^{-3}$ ). Figure  \ref{butane} shows the ion signal as a function of the photon energy (10.3 to 12 eV, 10 meV step) acquired over an hour (i.e. 170 steps averaged over 20 s per step). The ion signal at each photon energy was normalized for the DESIRS beamline photon flux, as measured by a calibrated, VUV-sensitive photodiode. Data obtained are in excellent agreement with those obtained by Wang et al.\citep{Wang_2008} The appearance energy for n-butane ($m/z$=58) is consistent with its ionization energy (10.53 eV) while fragment ions,  C$_3$H$_6
^+$ ($m/z$=42) and  C$_3$H$_7^+$ ($m/z$=43) are observed for photon energies of ca. 11.05 eV, i.e. only  0.5 eV above the molecular ionization threshold of butane, illustrating the  importance of using near-threshold ionization techniques to identify molecular species formed by collisions without suffering from unwanted fragmentation effects. 

\subsection{Formic acid dimer detection}

 Nucleation phenomena are of strong interest for planetary atmospheres in particular to model the sources and processes contributing to formation and growth of nanoparticles.\citep{zhang_nucleation_2012}  Kinetic studies of homogeneous nucleation of small molecular species can provide valuable information on low temperature energy transfer, collision rates, and binding energies of van der Waals bound molecules.\citep{vehkamaki_thermodynamics_2012, BourgalaisPRL20016}
 

Formic acid is the simplest organic acid and represents a prototype for more complex carboxylic acids used by life on Earth. It has also been observed recently in planetary disks.\citep{Favre_2018} The formic acid dimer (HCOOH)$_2$, an eight-membered ring with double hydrogen bonds in its most stable configuration, has been extensively studied as a prototype for doubly  hydrogen bonded complexes. \cite{Farfan_2017}
{The dissociation energy and the binding energy of the formic acid dimer were recently estimated theoretically  by Miliordos et al.} \citep{Miliordos_2017} {to be $D_0$  = 59.77  $\pm$ 0.4 kJ/mol and $D_e$  = 67.30 $\pm$ 0.4 kJ/mol respectively, in excellent agreement with the binding energy estimated by Kollipost et al.}\citep{Kollipost_2012} {from their experimental results ($D_0$ = 59.5 $\pm$ 0.5 kJ/mol).}

\begin{figure}[h]
\includegraphics[width=\columnwidth]{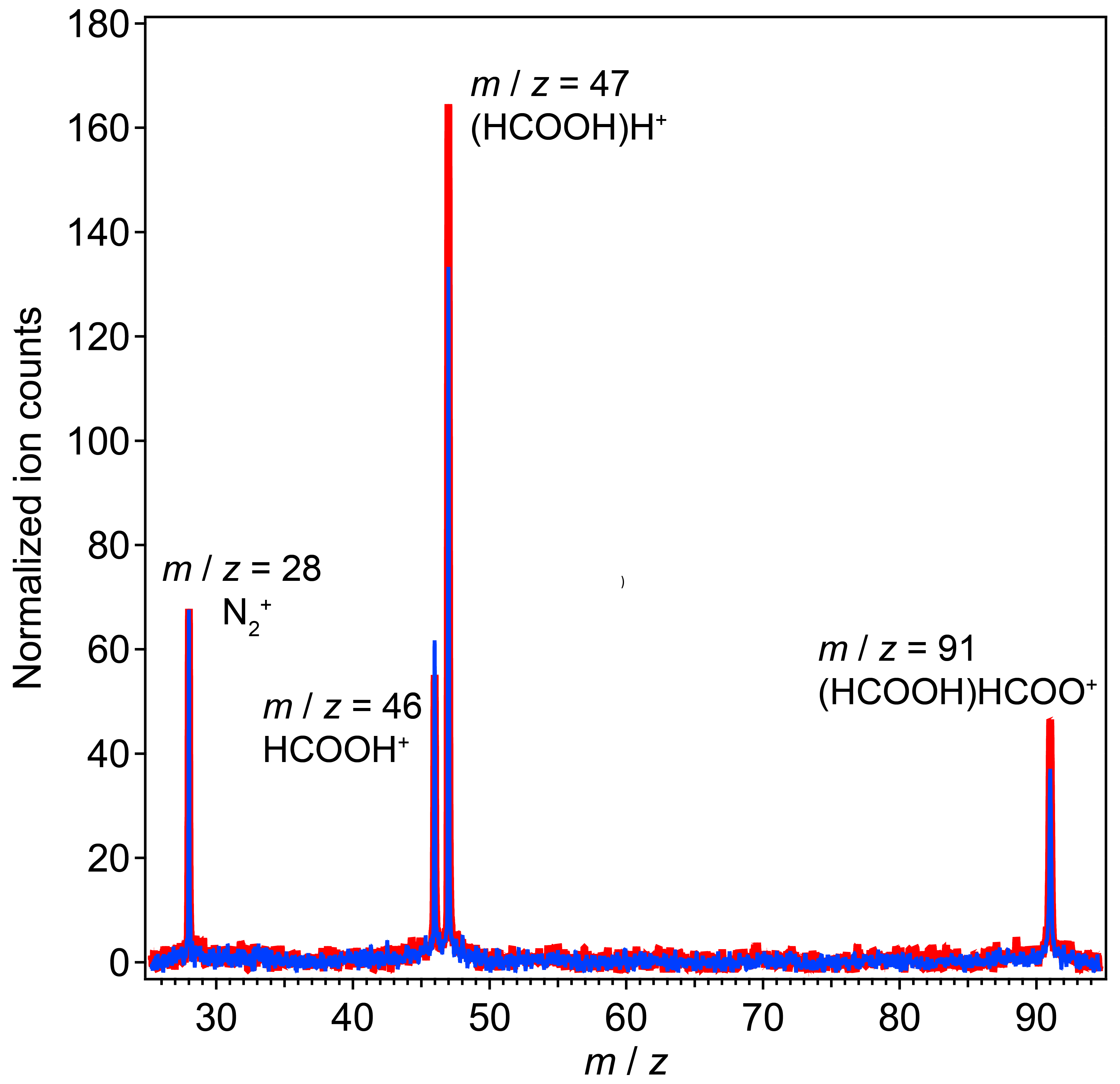}
 \caption{PEPICO mass spectra of a N$_2$/HCOOH flow at 74.6 K ($n_{\text{HCOOH}}$ = 2.37 $\times$ 10$^{14}$ cm$^{-3}$, total density = 6.6 $\times 10^{16}$~cm$^{-3}$) recorded at a photon energy of 11.28 eV, accumulated for 200 s. The blue spectrum was recorded at a distance from the end of the Laval nozzle to the skimmer of 2.5 cm, corresponding to a flow time of 19 $\mu$s, while the red spectrum was recorded at 25 cm, corresponding to a time of 193 $\mu$s.}
 \label{Formic_Acid_dimer}
\end{figure}

To observe formic acid dimers in our experiment, a small fraction (less than 1 \%) of formic acid was introduced in the buffer gas (N$_2$) of the CRESU flow. The dimer was first observed at room temperature (in subsonic conditions), then at 74.6 and 52 K using Laval nozzles whose flow properties are shown in Table 1.  Figure \ref{Formic_Acid_dimer}  displays mass spectra obtained by introducing a density of $n_{\text{HCOOH}}$ = 2.37 $\times$ 10$^{14}$ cm$^{-3}$ in the supersonic flow of a Laval nozzle operating at 74.6 K with N$_2$ as the buffer gas, for a total flow density of $n = 6.6  \times$ 10$^{16}$ cm$^{-3}$. 
The data were recorded at 11.28 eV photon energy accumulated for 200 s. Ion signals are observed at $m/z$ = 28, 46, 47 and 91. No signal was observed above $m/z$ = 91. The ion signal at $m / z$ = 46 corresponds to (HCOOH)$^+$, while $m / z$ at 47 and 91 are identified as fragments of formic acid dimers, (HCOOH)H$^+$ and (HCOOH)(HCOO)$^+$ respectively. The signal at $m / z$ = 28 is due to the ionization of the buffer gas, N$_2$, by higher energy photons from unfiltered harmonics of the undulator.

Formic acid monomers and dimers formed in the supersonic expansion can be monitored by skimming the uniform supersonic flow at various distances from the exit of the Laval nozzle that can be converted into reaction times. Accordingly, Figure \ref{Formic_Acid_dimer} shows two mass spectra taken at different distances corresponding to different reaction times: the blue curve at 2.5 cm, 19 $\mu$s, and the red curve at 25 cm, 193 $\mu$s. A reduction in the monomer signal at $m / z$ = 46 and an increase in the dimer peaks at $m / z$ = 47 and 91 can clearly be seen in passing from the short-time blue spectrum to the longer-time red spectrum, demonstrating the ability of the CRESUSOL experiment to follow the formation kinetics of weakly-bound clusters in cold supersonic flows. The kinetic data analysis for experiments performed under various conditions, including various densities of formic acid and different ionization energies, is in progress with the aim of determining rate coefficients for the dimerization of formic acid at 74.6 and 52.0 K.\citep{durif2020}

\subsection{Detection of C$_4$H$_2$ produced by the reaction between C$_2$H and C$_2$H$_2$ at low temperature}

The reaction C$_2$H + C$_2$H$_2 \rightarrow  $ C$_4$H$_2$+H was chosen for the first attempt to detect a bimolecular reaction product, here diacetylene, with the CRESUSOL apparatus at very low temperatures, 74.6 and 52.0 K.  

Ethynyl radicals, C$_2$H, have been detected both in the interstellar medium and in planetary atmospheres where they are known to participate in rapid reactions with hydrocarbons leading to molecular-weight growth with formation of carbon chains or polycyclic aromatic hydrocarbons (PAHs) \citep{Pedersen_1993, Opansky_1996, Hoobler_1999, Lee_2000, Carty_2001, 
Vakhtin_2001, Woon_2009, Soorkia_2010, Soorkia_2011, Landera_2011, Lockyear_2015}. 
Unsaturated hydrocarbons such as acetylene, C$_2$H$_2$, and diacetylene, C$_4$H$_2$, are also significantly abundant in many astrophysical objects where they are thought to be key intermediates for the production of larger species. \citep{Pedersen_1993,Lee_2000,Soorkia_2011,Sun_2015}
In particular, the reaction between C$_2$H and C$_2$H$_2$, is expected to proceed via addition-elimination to form C$_4$H$_2$ and H, even at low temperature, as shown by Soorkia et al.\citep{Soorkia_2011} 

In the present experiment C$_2$H radicals are generated by pulsed (50 Hz) laser photolysis of C$_2$H$_2$  using the 193 nm radiation of an ArF excimer laser (Lambda Physics COMPex-Pro 201). The laser beam enters the CRESU chamber and the reservoir through quartz windows (see Fig. 2) and propagates along the length of the CRESU flow. The fluence per pulse at the exit of the nozzle was measured to be 33 mJ cm$^{-2}$ in the absence of gas flow.

Acetylene (C$_2$H$_2$) was introduced in CRESU flows generated by Laval nozzles operating at 52.0 K and 74.6 K, with a density of typically 10$^{15}$ cm$^{-3}$. Given the absorption cross section of C$_2$H$_2$ at 193 nm \citep{Seki_1993}, $\sigma = 1.4 \times 10^{-19}$ cm$^{2}$  and assuming a quantum yield\citep{Okabe_1983} of 0.3, the typical C$_2$H density formed in the CRESU flow was therefore estimated to be ca. 10$^{12}$ cm$^{-3}$. Given that the reaction of C$_2$H with C$_2$H$_2$ is a single channel reaction, in our experimental conditions the maximum C$_4$H$_2$ density produced by the reaction cannot be higher than the density of C$_2$H estimated above. 

\begin{figure}[t]
 \includegraphics[width=\columnwidth]{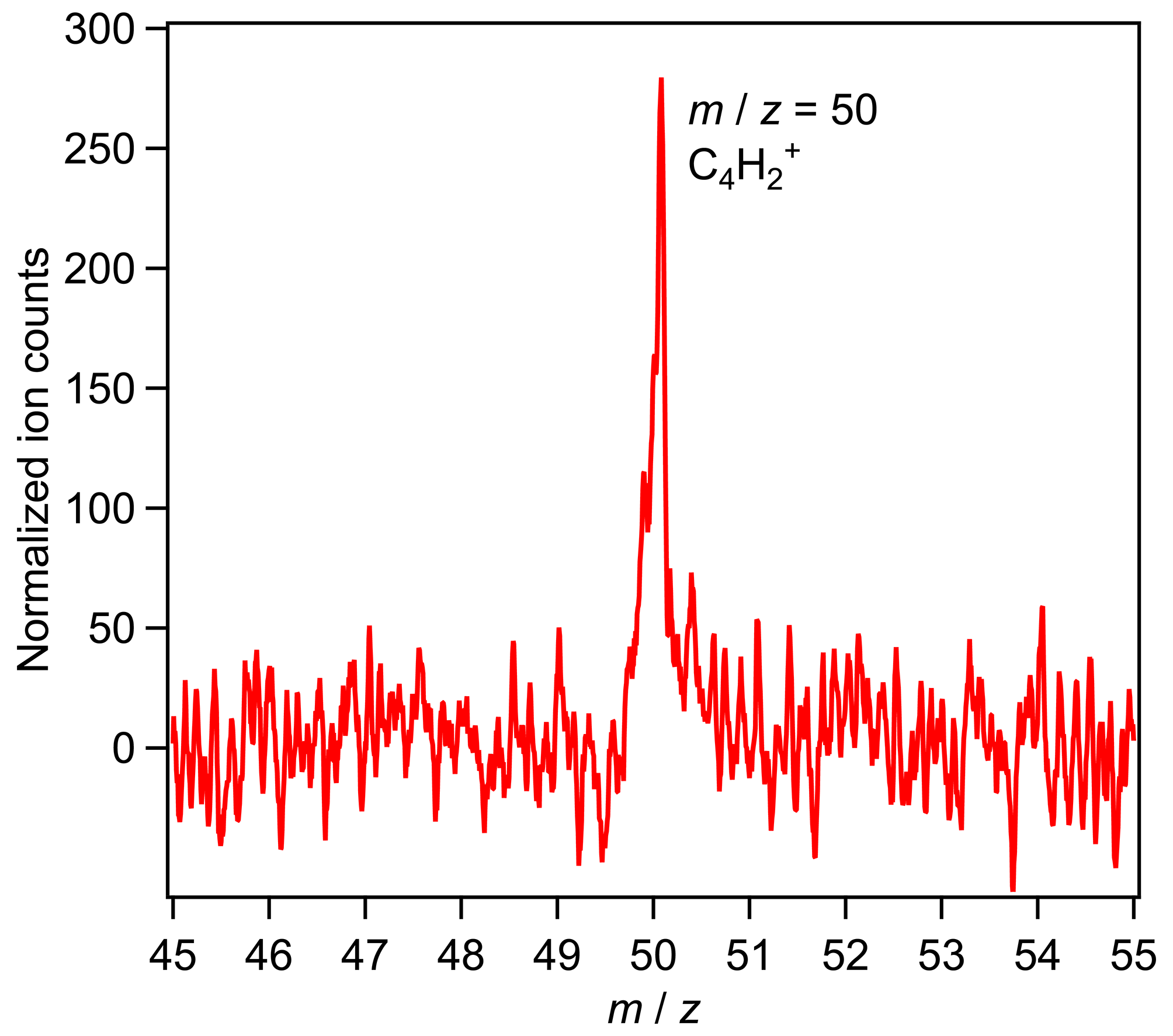}
\caption{Mass spectrum of C$_4$H$_2$ reaction product obtained from photolysis of C$_2$H$_2$  ($n_{\text{C}_2\text{H}_2}$ = 1.8 $\times 10^{15}$~cm$^{-3}$) introduced in a 74.6 K uniform supersonic flow of nitrogen (total density = 6.6 $\times 10^{16}$~cm$^{-3}$). {See text for full details}.}
 \label{C4H2}
\end{figure}

Figure \ref{C4H2} displays the mass spectrum obtained by photolysis of C$_2$H$_2$ ($n_{\text{C}_2\text{H}_2}$ = 1.6 $\times$ 10$^{15}$ cm$^{-3}$) introduced in the supersonic flow at 74.6 K with nitrogen used as buffer gas (total density = 6.6 $\times 10^{16}$~cm$^{-3}$). The data are integrated over a photon energy range between 10.0 eV to 11.35 eV, $\Delta E = 0.05$ eV, with an acquisition time of 500 s per energy step. This range spans the ionization energy of the expected product C$_4$H$_2$ (I.E. = 10.17 eV) and stops just below the ionization energy of C$_2$H$_2$ (11.40 eV). {The data are integrated from 0 to 600 $\mu$s after the laser shot, corresponding to the duration of the cold uniform supersonic flow. The false coincidence background has also been removed.} The signal at $m$ / $z$ = 50 shows the presence of C$_4$H$_2$ formed by the reaction between C$_2$H and C$_2$H$_2$, which is known to be very fast down to 15 K with rate coefficients greater than 10$^{-10}$ cm$^3$ s$^{-1}$ at any temperature below 100 K.\citep{Chastaing_1998} {Due to the relatively high repetition rate of the excimer laser (50 Hz), some residual C$_4$H$_2$ created by photolysis and reaction from preceding laser shots remains in the reservoir. In order to show only C$_4$H$_2$ produced in the cold supersonic flow, an equivalent  mass  spectrum  integrated over the same length of time (600 $\mu$s) but taken from 7.2 to 7.8 ms after the laser shot has been subtracted from the data, and a Savitzky-Golay fourth order 23 point smooth has been applied to give the mass spectrum shown in Figure \ref{C4H2}.}


\section{Conclusion and perspectives}

In this paper, we have presented a new instrument, CRESUSOL, dedicated to the kinetic study of low-temperature gas phase neutral-neutral reactions including clustering processes. This apparatus  couples tunable VUV synchrotron photoionization and time of flight mass spectrometry. Reactants,  and their subsequent products are probed by a modified Wiley-McLaren time-of-flight mass spectrometer collecting ions and triggered by electrons detected in coincidence. The first results shown here demonstrate the potential of the apparatus and reveal some limits that can be overcome by further developments. 

In its present configuration, the results obtained on the formation of formic acid dimers show that the CRESUSOL apparatus can monitor both the kinetic evolution of the dimer and the monomer channels, and will give access to the rate coefficient of dimerization of formic acid for the first time at low temperature. \citep{durif2020} 
The tunability offered by the synchrotron radiation is particularly important for such studies as it allows a very soft ionization of the clusters which limits their evaporation and fragmentation. The study of cluster formation under well controlled conditions of temperature and pressure, is of great importance to understand the first steps of formation of particles at the molecular level. Such kinetic studies, combined with theoretical calculations, will bring valuable information on important physical parameters like energy transfer and binding energies of van der Waals bound molecules at low temperatures. 

This work has also shown the need to undertake some improvements of the current configuration of the CRESUSOL apparatus. The large number of false coincidences reduces significantly the sensitivity of the detection and precludes the identification of reaction products of chemical reactions in the appropriate conditions for performing kinetic measurement in CRESU flows. The density of reactants that can be introduced in supersonic flows indeed, cannot exceed a few percents of the total gas density (typically $10^{16}-10^{17} $cm$^{-3}$) in order to preserve the uniformity of the flows and subsequently the temperature conditions expected. The reaction between C$_2$H and C$_2$H$_2$ is a particularly favourable case because (i) the co-reactant C$_2$H$_2$ is also the precursor of radical  C$_2$H, (ii) the ionization energy of the reactants, 11.61 and 11.40 eV  for C$_2$H and C$_2$H$_2$ respectively, are higher than the ionization energy of the expected product, C$_4$H$_2$, 10.17 eV. In order to improve substantially the sensitivity of the apparatus, an update of the mass spectrometer will be introduced in the medium-term.  A position sensitive detector (PSD) will be implemented on the ion detector side of the time-of flight mass spectrometer. This is expected to increase by orders of magnitude the ion signal contrast {minimising} the contribution {of false coincidences and} background species present in the chamber. {Indeed, the species from the CRESU flow have a large velocity along the flow direction and therefore arrive at a position of the detector which depends on $\sqrt{m/z}$. For a false coincidence to be counted not only has it to arrive within the same time range as a true coincidence, but also at the correct position, which dramatically decreases the probability of it occurring. In addition, the background gas velocity along the flow has a thermal distribution which is centered at zero, so that the arrival position of background species will be markedly different from that of the CRESU species, facilitating their removal and increasing the signal-to-background.}

In the near future it is also planned to implement a temporal ion deflection coupled with a position-sensitive ion detector as proposed by Osborn and co-workers.\citep{Osborn_2016}
This device should largely remove the false coincidence background, increasing the dynamic range in the PEPICO TOF mass spectrum by 2–3 orders of magnitude. 

In the longer term, pulsed uniform supersonic flows will be implemented in the CRESUSOL apparatus allowing it to reach temperatures much lower than 50 K, in a temperature domain where quantum effects such as tunneling prevail. New findings can be expected that will be of interest for both fundamental physical chemistry and astrochemistry as a whole, as it will certainly show new formation channels for complex molecules in these particular and extreme temperature conditions 
typical of star and planet forming regions.

\section*{Acknowledgments}

The authors would like to thank Jacques Sorieux, Didier Biet, Ewen Gallou, Guy Pécheul, Yvonig Robert and Alexandre Dapp from the Université de Rennes 1 mechanical workshop for their technical support, as well as Gregory H. Jones from Caltech for assistance in performing some of the experiments. The authors also thank Bertrand Rowe and Sébastien Moralès for valuable scientific discussions ahead of the project implementation. This work is supported by the French National Research Agency through the project CRESUSOL (ANR-11-BS04-0024) and the CNRS-INSU 'Programme National de Physique et Chimie du Milieu Interstellaire (PCMI)' and 'Programme National de Plan\'etologie (PNP)'. SDLP acknowledges financial support from the Institut Universitaire de France. This work was also supported in part by the National Science Foundation under Grant No. CHE-1413712. JPM was supported by the National Science Foundation Graduate Research Fellowship (NSF GRFP) and the National Science Foundation Graduate Research Opportunities Worldwide (NSF GROW) programs. JPM would also like to thank the Office for Science and Technology of the Embassy of France in the United States for a Chateaubriand Fellowship. The authors thank the QUADMARTS International Research Network for promoting their collaboration. We warmly thank the whole SOLEIL staff for providing beamtime under projects \#20160297 and \#20181883.
\section*{DATA AVAILABILITY}
The data that support the findings of this study are available from the corresponding author upon reasonable request.
\section*{References}
\bibliography{CRESUSOL_Refs}

\end{document}